\documentclass[twocolumn,superscriptaddress,floatfix]{revtex4}
\usepackage[utf8]{inputenc}
\usepackage{color}
\usepackage{graphicx}  
\usepackage{psfrag}
\usepackage{bm}
\usepackage{wasysym}
\usepackage{units}

\begin{document}
\title{Slicing the $3d$ Ising model: critical equilibrium and coarsening dynamics} 


\author{Jeferson J. Arenzon}
\affiliation{Instituto de F\'\i sica, Universidade Federal do Rio Grande do Sul,\\ CP 15051, 91501-970 Porto Alegre RS, Brazil} 

\author{Leticia F. Cugliandolo}
\affiliation{Sorbonne Universit\'es, Universit\'e Pierre et Marie Curie -- Paris VI,\\ Laboratoire de Physique Th\'eorique et Hautes 
Energies UMR 7589,
 \\ 4 Place Jussieu, 75252 Paris Cedex 05, France} 

\author{Marco Picco}
\affiliation{Sorbonne Universit\'es, Universit\'e Pierre et Marie Curie -- Paris VI,\\ Laboratoire de Physique Th\'eorique et Hautes 
Energies UMR 7589,
 \\ 4 Place Jussieu, 75252 Paris Cedex 05, France} 

\vspace{0.25cm}

\date{\today}  
\begin{abstract}
We study the evolution of spin clusters on two 
dimensional slices of the $3d$ Ising model in contact with a heat bath after
a sudden quench to a subcritical temperature. 
We analyze the evolution of some simple initial configurations,  such as a sphere and a torus, of one 
phase embedded into the other, to confirm that their area disappears linearly in time and to establish the 
temperature dependence of the prefactor in each case.
Two generic kinds of initial states are later used: equilibrium configurations either
at infinite temperature or at the paramagnetic-ferromagnetic phase transition. We 
investigate the morphological 
domain structure of the coarsening configurations on $2d$ slices of the $3d$ system,
comparing with the behavior of the bidimensional model.
\end{abstract}
\maketitle

\section{Introduction}

Phase ordering kinetics is a phenomenon  often encountered in nature. Systems 
with this kind of dynamics
provide, possibly, the simplest realization of cooperative out of equilibrium dynamics at macroscopic scales. For such systems, 
the mechanisms whereby the relaxation takes place are usually well-understood~\cite{Bray94, Onuki02, PuWa09}, but quantitative predictions on 
relevant observables are
hard to derive analytically. Coarsening systems are important from a fundamental point of view as they pose 
many technical questions that are also encountered in other macroscopic systems out of equilibrium 
that are not so well-understood, such as glasses or active matter. They are also important from the standpoint of 
applications as the macroscopic properties of many materials depend upon their domain morphology.

The hallmark of coarsening systems is dynamic scaling, that is to say, the fact that the 
morphological pattern of domains at earlier times looks statistically similar to the
pattern at later times apart from a global change of scale~\cite{Bray94, Onuki02, PuWa09}. Dynamic scaling has been successfully 
used to describe the dynamic structure factor measured with scattering methods, 
and the space-time correlations computed numerically in many models. It has  
also been shown in a few exactly solvable 
cases and within analytic approximations to coarse-grained models. 

New experimental techniques make now possible the direct visualization of the domain 
structure of three-dimensional coarsening systems. In earlier studies, the 
domain structure was usually observed {\it post mortem}, and only on exposed two-dimensional slices of the 
samples, with optic or electronic microscopy. 
Nowadays, it became possible to observe the full $3d$ micro structure {\it in situ} and  in the course of evolution.
These methods open the way to observe microscopic processes that were so 
far out of experimental reach. 
For instance, in the context of  soft-matter systems, 
laser scanning confocal microscopy 
was applied to  phase separating binary liquids~\cite{WhiteWiltzius95} and polymer blends~\cite{Jinnai95,Jinnai97,Jinnai99},
while X-ray tomography was used to observe phase separating glass-forming liquid mixtures~\cite{Bouttes}
and the time evolution of foams towards the scaling state~\cite{Lambert2010}. 
In the realm of magnetic systems the method presented in~\cite{Manke-etal10} looks very promising.

Three dimensional images give, in principle, access to a complete topological characterization of the 
interfaces via the calculation of quantities such as the Euler characteristics and the local mean and Gaussian curvatures. 
On top of these very detailed analyses, one can also extract the evolution 
of the morphological domain structure on different planes across the samples and investigate 
up to which extent the third dimension has an effect on what occurs in strictly two dimensions. 

In this paper the focus is set on the dynamic universality class of non-conserved scalar order parameter,
as realized by Ising-like magnetic samples taken into the ferromagnetic phase across their second order 
phase transition. An important question is to which extent the results for the morphological properties of 
strictly $2d$ coarsening apply to the $2d$ slices of $3d$ coarsening.
In the theoretical study of phase ordering kinetics, a continuous coarse-grained description of 
the domain growth process, in the form of a time-dependent Ginzburg-Landau equation, is used.  
Within this approach, at zero temperature, 
the local velocity of any interface is proportional to its local mean curvature. Accordingly, in $2d$
the domains can neither merge nor disconnect in two (or more) components. 
In $2d$ one can further exploit the fact that the dynamics are curvature driven and use the Gauss-Bonnet theorem 
to find an approximate expressions for several statistical and geometric properties that characterize the domain 
structure. In this way, expressions for the number density of domain areas, number density of perimeter lengths,
the relation between the area and the length of a domain, etc. were found~\cite{ArBrCuSi07,SiArBrCu07}.
 In $3d$, instead, the curvature driven dynamics do not prohibit breaking a domain in two or merging two 
domains, and the Gauss-Bonnet theorem, that involves the Gaussian curvature instead of the mean curvature,
cannot be used to derive expressions for the statistical and geometric properties of volumes and areas.
Moreover, merging can occur in two-dimensional slices of a three-dimensional system, for example, {\it via} 
the escape of the opposite phase in-between into the perpendicular direction to the plane.

Two previous studies of $3d$ domain growth are worth mentioning here, although their 
focus was different from ours as we explain below.

The morphology of the $3d$ zero-temperature non-conserved scalar 
order parameter coarsening was addressed in~\cite{Holyst01,Holyst02}. 
From the numerical solution of the time-dependent Ginzburg-Landau equation,
results on the time-dependence of the topological properties  of the interfaces 
were obtained. 

The late time dynamics of the $3d$ Ising model evolving at zero temperature 
was analyzed in~\cite{SpKrRe01,SpKrRe02,OlKrRe11a,OlKrRe11b}. 
It was shown in these papers that  the $3d$IM does not reach the 
ground state nor a frozen state at vanishing temperature. Instead, it stays wandering around an iso-energy subspace of phase space
made of metastable states that differ from one another by the state of 
blinking spins that flip at no energy cost~\cite{SpKrRe01}. At very low temperature the relaxation proceeds
in two steps. First, with the formation of a metastable state similar to the ones of zero-temperature; 
next, with the actual approach to equilibrium~\cite{SpKrRe02}. The sponge-like nature of the 
metastable states was  examined in~\cite{SpKrRe02,OlKrRe11a,OlKrRe11b}.

In our study we use Monte Carlo simulations of 
the  $3d$ Ising Model (IM) on a cubic lattice with periodic boundary conditions, and we focus
on the statistical and geometrical properties of the geometric domains and hull-enclosed areas on planes 
of the cubic lattice. A number of equilibrium 
critical properties, necessary to better understand our study of the coarsening dynamics in
Sec.~\ref{sec.coarsening}, are first revisited in Sec.~\ref{sec.eq}. We start
the study of the dynamics by comparing the contraction of a
spherical domain immersed in the background of the opposite phase in 
$d=2$ and $d=3$, both at zero and finite temperature. This study, even
though for a symmetric and isolated domain, allows us to evaluate the dynamic 
growing length and its temperature dependence. When the domains are not isolated,
as is the case, for instance, of two circular slices lying on the same plane but being associated to the same $3d$ torus,
merging may occur along evolution, and this process contributes to the complexity of the problem. 
 We also study the dynamic scaling of the 
space-time correlation on the $2d$ slices. Following these introductory parts,  
the statistical and morphological properties of the areas and perimeters of geometric domains and hull-enclosed areas 
on $2d$ slices of the $3d$IM are then presented. We end by summarising our results and by discussing 
some lines for future research in Sec.~\ref{sec:conclusions}.

\section{The model and its equilibrium properties}
\label{sec.eq}

Before approaching the dynamic problem we need to define the model and establish some of its equilibrium 
properties. This is the purpose of this section. The system sizes used in the 
equilibrium simulations range from $L=40$  to $L=800$ in $d=2$ and from $L=40$ to $L=400$ in $d=3$. 
The samples at the critical point were equilibrated with the usual cluster 
algorithms~\cite{NeBa99}. 

\subsection{The model}

The Ising model (IM)
\begin{equation}
H_J = - J\sum_{\langle ij\rangle} s_i s_j 
\end{equation}
with $s_i \pm 1$,  $J>0$, and the sum running over nearest neighbors on a $d>1$ lattice, 
undergoes an equilibrium  second order phase transition  at the Curie 
temperature $T_c>0$. The upper critical phase is paramagnetic and the lower critical 
phase is ferromagnetic. 
 
In $2d$ the  critical temperature $T_c$ coincides with the 
temperature at which the geometric clusters (a set of nearest neighbor equally
oriented spins) of the two phases percolate~\cite{Binder76,CoNaPeRu77}. 
 This is not the case in $3d$: the percolation temperature, $T_p$, at which 
a geometric cluster of the minority phase percolates, is lower than the Curie temperature $T_c$~\cite{Muller74}. 
On the cubic lattice, that we will use in this work, $T_c \simeq 4.5115$~\cite{TaBl96} and $T_p=0.92 \ T_c$~\cite{Muller74}. 
The random site percolation threshold on the cubic lattice is $p_c\simeq 0.312$~\cite{Stauffer94}.
The boundary conditions in the simulations are periodic.

\subsection{Equilibrium domain area distribution at $T_c$}

As already stated, the spin clusters in the $3d$IM {\it are not} critical at the 
magnetic second-order phase transition~\cite{Muller74}. Still, the $2d$ spin clusters 
on a slice, defined by taking all the spins in a $3d$ lattice $(x,y,z)$ with either $x$, 
$y$ or $z$ fixed, {\it are} critical with properties of a new universality 
class~\cite{DoPiWiHaMaMa95}. 
In particular, the distribution of the length of the surrounding interfaces, $N(\ell)$, and the area as a function of this 
length, $A(\ell)$, satisfy 
\begin{equation}
N(\ell) \simeq \ell^{-\tau_{\ell}} \; ,  \quad \qquad A(\ell) \simeq \ell^{\delta} \; ,
\end{equation}
with $\tau_{\ell} \simeq 2.23 (1)$ and $\delta \simeq 1.23(1)$.
Similar measurements were performed in Ref.~\cite{SaDa10} 
where a consistent value of the fractal dimension $d_f = 2/\delta$ was obtained. 
For the sake of comparison, the values of the exponents $\tau_{\ell}$ and $\delta$ for 
the critical spin clusters in $2d$~\cite{CaNa86,VaSt89,SiArBrCu07} are $\tau_{\ell}^{(2d)}=27/11\simeq 2.45455$ 
and $\delta^{(2d)} = 16/11 \simeq 1.45455 = \tau_{\ell}^{(2d)} -1 $.

As the main purpose of this paper is to characterize the out of equilibrium dynamics
of the $3d$IM by focusing on the behavior of the statistical and geometrical properties of the 
cluster areas on $2d$ slices, we  further investigated the equilibrium properties of the 
same objects at the phase transition in $2d$ and $3d$, with the purpose of checking whether the 
theoretical expectations are realized numerically for the system sizes we can simulate. 

We consider the number density of cluster areas on a two dimensional plane. The area is defined
as the number of spins in a geometric cluster that are 
connected (as first neighbors on the square 
lattice) to its border. With this definition one excludes the spins in the holes 
inside a cluster, that is to say, one considers the proper domain areas. At criticality,
the number density of {\it finite areas}, 
excluding the percolating cluster, is given by the power law 
\begin{equation}
N(A) \simeq A^{-\tau_A} \; .
\end{equation}
The contribution of the percolating clusters can be included as
an extra term that takes into account that these clusters 
should scale as $L^{2-(\beta/\nu)_s}$ where $(\beta/\nu)_s$ is the ``magnetic 
exponent" associated to the spin clusters~\cite{Stauffer94}. Thus, for a system with 
linear size $L$, the  {\it full distribution} is
\begin{equation}
{\cal N}(A) = L^2 N(A) + a \delta(A - b L^{2-(\beta/\nu)_s}) 
\label{eq.fulldistribution}
\end{equation}
and, by construction,  the normalization condition is 
\begin{equation}
\label{norm}
\int dA \ A \ {\cal N}(A) = L^2 \; .
\end{equation}
If the distribution $N(A)$ has a bounded support, with a maximum cluster of size $M_L$, then 
\begin{eqnarray}
\int dA \ A \ {\cal N}(A)  &\simeq& \frac{L^2}{\tau_A -2} (A_0^{2-\tau_A} 
                           - M_L^{2-\tau_A }) \nonumber\\ 
                           &+& a b L^{2-(\beta/\nu)_s} 
\;\;\;,
\end{eqnarray}
where $A_0$ is a microscopic area. The normalization condition can be satisfied only if  
\begin{equation}
M_L \simeq L^{-{(\beta/\nu)_s \over 2-\tau_A }} 
\label{eq.ML1}
\end{equation}
that, using a standard relation between exponents from percolation theory~\cite{Stauffer94},
\begin{equation}
\label{bt}
(\beta/\nu)_s = d \left({\tau_A -2 \over \tau_A-1} \right)
\; , 
\end{equation}
yields
\begin{equation}
\label{ML}
M_L \simeq L^{2/(\tau_A-1)} \simeq L^{2 - (\beta/\nu)_s}
\; .
\label{eq.ML2}
\end{equation}
This implies that the largest size contributing to the distribution $N(A)$ has
the same fractal dimension of the percolating cluster, thus scaling with the same 
power as the ``magnetic term". The distribution of finite areas extends its support to 
a size-dependent size such that it matches the weight of the percolating 
clusters. In other words,  there is no gap between the two contributions
to Eq.~(\ref{eq.fulldistribution}). 
In the following, we will first examine these relations in the $2d$IM at its 
critical point and
we will come back to the slicing of the $3d$ systems later. 

\subsubsection{The critical $2d$IM}

\begin{figure}
\includegraphics[width=8cm]{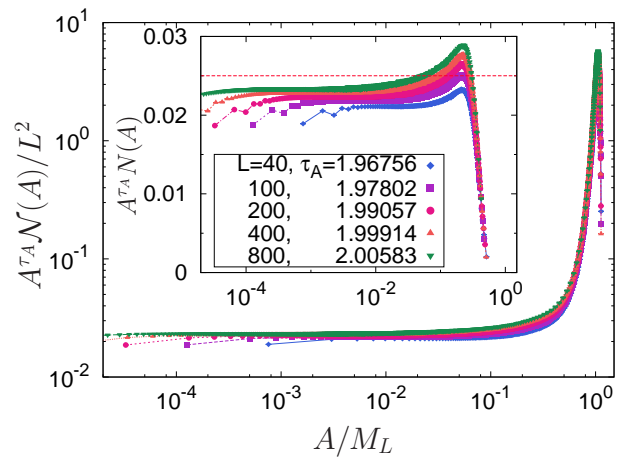}
\caption{(Color online.) 
Distribution of the spin cluster sizes in the $2d$IM at its critical temperature. 
Scaling of the number density of finite areas, $A^{\tau_A} N(A)$ (inset), and
the full distribution of all areas (main panel) {\it vs.} $A/M_L$ with $M_L$ given by Eq.~(\ref{eq.ML}), 
for various system sizes $L$ given in the key together with the values of $\tau_A$ used. The
dashed horizontal line in the inset is $c_d\simeq 0.025$~\cite{SiArBrCu07}, to which the
plateau should asymptotically converge.
}
\label{F1}
\end{figure}

\begin{figure}
\includegraphics[width=8cm]{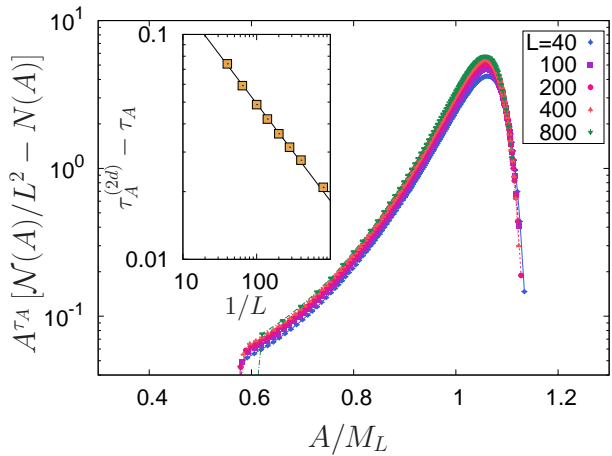}
\caption{(Color online.)
Distribution of the percolating spin cluster areas in the $2d$IM at its critical temperature. 
Scaling of the number density of percolating areas, $A^{\tau_A} [{\cal N}(A) /L^2-N(A)]$  {\it vs.} $A/M_L$ with $M_L$ given 
by Eq.~(\ref{eq.ML}), for various system sizes $L$ given in the key. Inset: values of $\tau_A$ from
Fig.~\ref{F1} as a function of $1/L$ showing the convergence to the asymptotic value  $\tau_A^{(2d)} = 379/187$.
}
\label{F2}
\end{figure}

We first check the predictions listed above in the $2d$IM  at its critical point. In this case
$(\beta/\nu)_s$ coincides with the magnetic exponent of the tricritical Potts model with $q=1$~\cite{StVa89}.
Therefore, 
\begin{equation}
(\beta/\nu)^{(2d)}_s = {5 \over 96} \; ,
\qquad\;\;\;
\tau_A^{(2d)} = \frac{379}{187}
\; , 
\end{equation}
implying
\begin{equation}
M_L  \simeq L^{187/96} \; .
\label{eq.ML}
\end{equation}
In Fig.~\ref{F1} (inset) we present $A^{\tau_A} N(A)$ against 
${A / M_L}$ with  $N(A)$ the distribution of finite spin clusters, {\it i.e.}  
excluding the percolating cluster from each configuration.
In this plot we used $M_L \simeq L^{2 - (\beta/\nu)^{(2d)}_s}$ with the exact value of $(\beta/\nu)^{(2d)}_s$, 
and we determined the value of $\tau^{(2d)}_A$ for each size $L$ finding that its 
dependence on $L$ is rather strong. It is only for the largest simulated system ($L=800$) that 
$\tau^{(2d)}_A$ becomes larger than 2, converging, in the thermodynamical limit, to the right value, $379/187$, as shown in the inset of Fig.~\ref{F2}.
Note, in the inset of Fig.~\ref{F1}, that $N(A)$ has a maximum for large clusters. These
clusters are non percolating: in the overwhelming majority of cases, if the largest cluster percolates, thus
contributing to the second term in ${\cal N}(A)$, the second largest does not and goes to $N(A)$.
With the above exponents we obtain a nice scaling of this maximum in $N(A)$ 
as well as of the peak in ${\cal N}(A)$, Fig.~\ref{F1} (main panel). This last feature can be magnified by
subtracting the support of finite clusters to leave only the areas of the
percolating clusters. Indeed, by plotting $A^{\tau_A} [{\cal N}(A) /L^2-N(A)]$ 
as a function of $A/M_L$, see Fig.~\ref{F2}, we obtain a perfect scaling. Moreover,
the plateau observed in the rescaled plot in the inset of Fig.~\ref{F1} (horizontal
dashed line), although
smaller than the estimated value in Ref.~\cite{SiArBrCu07}, $c_d\simeq 0.025$, when
extrapolated to very large sizes, is consistent with this value.

\subsubsection{Slicing the $3d$IM}

\begin{figure}
\includegraphics[width=8cm]{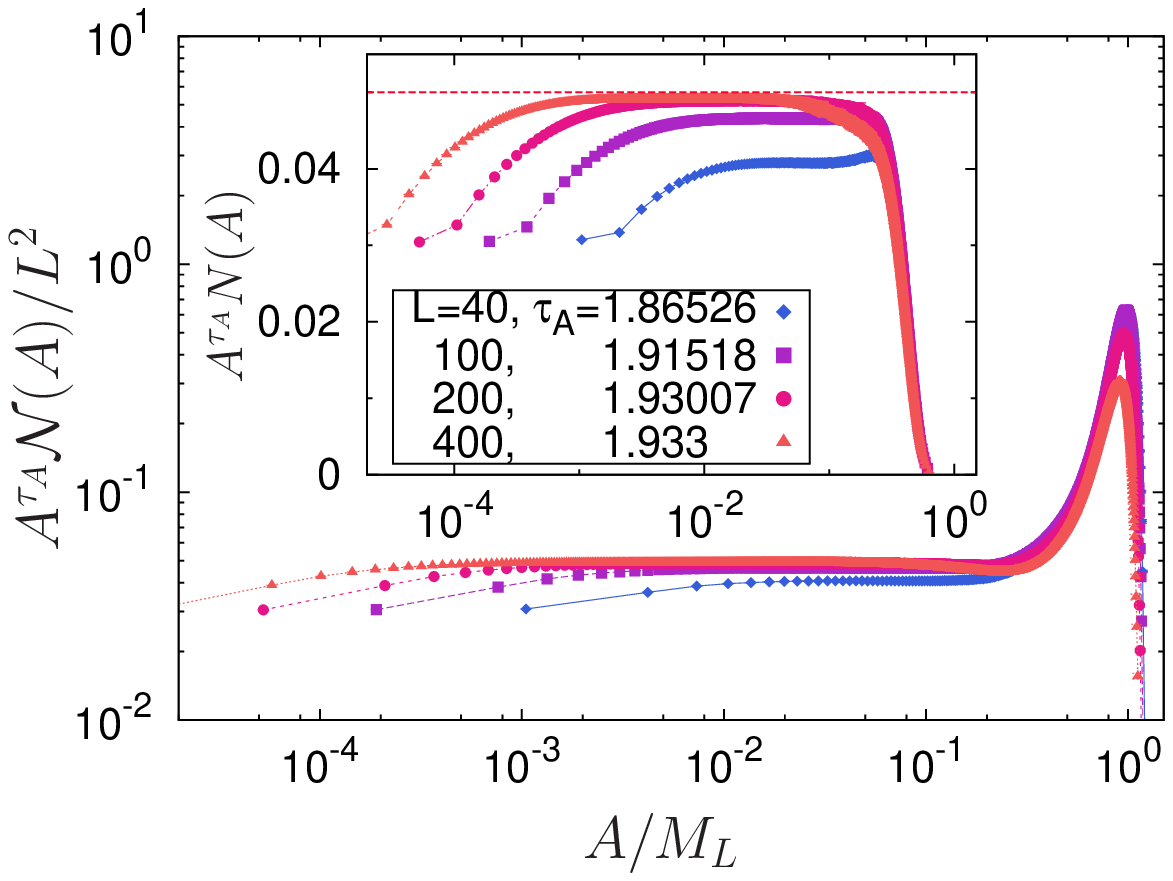}
\caption{(Color online.) Distribution of the spin cluster sizes for slices of the $3d$IM at its critical temperature. 
Scaling of the number density of finite areas, $A^{\tau_A} N(A)$ (inset), and
the full distribution of all areas (main panel) {\it vs.} $A/M_L$ with $M_L\simeq L^{1.86}$,
for various system sizes $L$ given in the key together with the values of $\tau_A$ used.
The dashed horizontal line in the inset is $2c_d\simeq 0.05$~\cite{SiArBrCu07}, twice the
value for the $2d$IM.
}
\label{F3}
\end{figure}

\begin{figure}
\includegraphics[width=8cm]{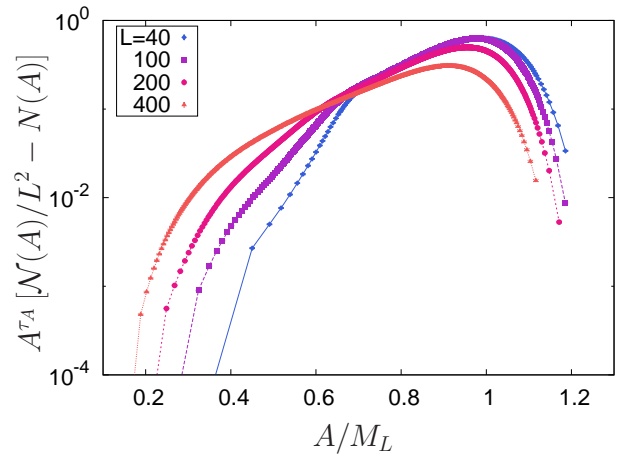}
\caption{(Color online.) 
Distribution of the size of the largest  cluster 
 $A^{\tau_A} \left[{\cal N}(A) /L^2 - N(A)\right] $ {\it vs.}  $A/M_L$ with $M_L\sim L^{1.86}$ for spin cluster areas  
 on slices of the $3d$IM at its critical temperature. Differently from
Fig.~\ref{F2}, the curves do not scale here.
}
\label{F4}
\end{figure}

Next we turn to the analysis of the domain areas on $2d$ slices of the critical $3d$IM. 
The values of  the exponents $(\beta/\nu)_s$ and $\tau_A$, and therefore the scaling of $M_L$ with $L$,
are not known for these objects and we study them here.
In Fig.~\ref{F3}~(inset) we show the distribution of the finite size spin clusters,  $N(A)$, and
we determine $\tau_A$ for each size $L$. The measured values, shown in the key, 
are much smaller than $2$ even for the largest simulated system and they seem 
to converge to a value $\tau_A \simeq 1.94 < 2$. This fact is clearly disturbing since it implies that 
$M_L$ would decrease to zero with increasing $L$.
In Fig.~\ref{F3} (main panel) we rescale $A$ as $A/L^{x}$ and we obtain a good collapse of data 
for large $A$ with $x=1.86$. This implies $M_L \simeq L^{1.86}$. 
In this figure, we also note that while the scaling of $N(A)$ is good for small values of the 
scaling variable, this is not the case for large ones. 
This fact is even clearer in Fig.~\ref{F4}
where we see that the part of the distribution that corresponds to the largest spin clusters does not scale.
We observed that  in some configurations the largest cluster does not percolate on the slice.
The lack of scaling in Fig.~\ref{F4} is probably related to the inconsistent value $\tau_A < 2$ that we obtained from the analysis of $N(A)$. 

There are two remarkable differences between $N(A)$ measured in the $2d$ system
and in the sliced $3d$ one. The first is that the height of the plateau (horizontal 
dashed line in the inset of Fig.~\ref{F3}) seems to converge to a value that is about  
twice the one found in the $2d$ case, that is, $2c_d\simeq 0.05$~\cite{SiArBrCu07}.
The second difference is that, in the slices, $N(A)$ does not present the maximum
observed in $2d$. This may explain why the distribution in this case is higher: the
absence of a second, large cluster creates a large amount of space that will be
filled by smaller ones. 


\begin{figure}
\includegraphics[width=8cm]{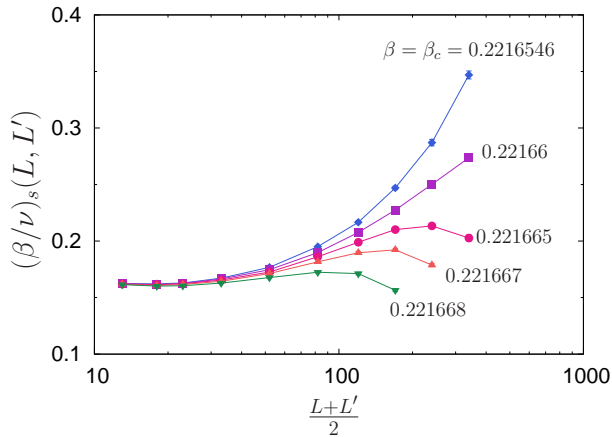}
\caption{(Color online.)
Effective value of $(\beta/\nu)_s(L,L')$ vs $(L+L')/2$ for the $3d$IM at its critical temperature 
as extracted from the analysis of the averaged size of the largest spin cluster on $2d$ slices, see Eq.~(\ref{eq:beta-s}).}
\label{F5}
\end{figure}

We thus conclude that the small size spin clusters living in slices of the $3d$IM scale 
at the critical point while the weight of the percolating clusters does not seem to do. 
In order to clarify this issue we used the expected scaling of the largest cluster with the system size, 
 $M_L \simeq L^{2 - (\beta/\nu)_s}$, Eq.~(\ref{ML}), to determine $(\beta/\nu)_s$.
 In Fig.~\ref{F5}, we show the values of the effective exponent $(\beta/\nu)_s$ obtained from a two points fit 
\begin{equation}
\left({\beta \over \nu}\right)_s(L,L') = 2-{\ln(M_L/M_{L'}) \over \ln(L/L')}
\; , 
\label{eq:beta-s}
\end{equation}
with $M_L$ the average size of the largest spin cluster. 
This quantity is expected to converge to a fixed value in the large size limit. However, 
for the sizes that we can simulate, it {\it does not} 
converge at the critical point. For the smallest sizes, $(\beta/\nu)_s(L,L')$ is 
close to a constant for $\beta = 1/T$ slightly larger than $\beta_c$.
Upon increasing the lattice size, we see that beyond some size, $(\beta/\nu)_s(L,L')$ drops. 
In the range $\beta_c \leq \beta < 0.221665$ we do not reach an asymptotic regime
and much larger sizes are needed to conclude on the actual value of $(\beta/\nu)_s$. 
This also means that the values of $\tau_A$ that we computed can still increase and eventually 
become larger than 2, as it should happen. It is interesting to point out that we checked 
that the same analysis done to the Fortuin-Kasteleyn  
clusters obtained from the same data yield a value of $(\beta/\nu)_s$ in perfect agreement with the 
theoretical expectation.

We conclude that it is very hard to reach the asymptotic, large size limit in which the 
values of the exponents $(\beta/\nu)_s$ and $\tau_A$ for the areas of the geometric clusters on $2d$ slices 
of the $3d$ system
should reach a stable limit.

\section{Coarsening properties}
\label{sec.coarsening}

Once the system is prepared (equilibrated) at a specific temperature, it will
be sub-critically quenched and the out of equilibrium subsequent dynamics,
studied. We start by presenting some background material on the $2d$ dynamics. We next 
describe the evolution of artificially designed single domain initial states (circular
or spherical in $2d$ or $3d$, respectively, and a torus). After having analyzed these simple situations,
the richer dynamics ensuing from an equilibrated state at $T_0\to\infty$ (non critical) and 
$T_0=T_c$ (critical on the slices) are studied.

\subsection{Background}
\label{sec.background}

With the coarse-grained approach, 
in two dimensions and in the absence of thermal fluctuations, one proves that the
number of hull-enclosed areas per unit area, $n_h(A,t)\,dA$, with enclosed area in
the interval $(A,A+dA)$, is~\cite{ArBrCuSi07,SiArBrCu07}
\begin{equation}
n_h(A,t)  =  \frac{(2)c_h}{(A+\lambda_{2d} t)^2}
\; ,
\label{eq.analytic-nh}
\end{equation}
where $c_h=1/(8\pi\sqrt{3})$ is a universal constant~\cite{CaZi03}. 
This result follows from the independent curvature driven evolution of the 
individual hull-enclosed areas from initial values taken from a 
probability distribution determined by the initial state of the system. The statistics
of the initial state is inherited in Eq.~(\ref{eq.analytic-nh}) by the factor in the numerator. Indeed, 
the factor 2 between parenthesis is present when the initial state is prepared at $T>T_c$. It
is due to the fact that the subcritical dynamics reach, after a time  that grows with the system 
size as $t_p \simeq L^{\alpha_p}$, critical percolation~\cite{BlCoCuPi14}. Instead, it is absent if the 
initial state is one of the critical Ising point. 

Temperature fluctuations have a double effect. On the one hand their effect is incorporated
in the factor $\lambda_{2d}$ that becomes $\lambda_{2d}(T)$ and takes into account the 
modification of the typical growing length (see below). On the other hand, small clusters
are created by these fluctuations and the 
distribution Eq.~(\ref{eq.analytic-nh}) has to be complemented with an exponentially decaying term that 
takes into account the additional weight of thermal equilibrium domains.

The number density of the areas of the geometric domains cannot be derived exactly. 
Under some reasonable assumptions, one argues~\cite{SiArBrCu07} that at zero working temperature
\begin{equation}
n_d(A,t)  =  \frac{(2)c_d (\lambda_{2d} t)^{\tau-2}}{(A+\lambda_{2d} t)^\tau}
\; ,
\label{eq.analytic-nd}
\end{equation}
with the constant $c_d\simeq 0.025$ being very close albeit different from $c_h$, 
and $\tau$ an exponent that takes the critical percolation or the critical Ising value  
depending on whether the initial state is a high temperature or a critical one.

The time dependence of these two number densities complies with dynamic scaling~\cite{Bray94}, 
with the typical length scaling as 
\begin{equation}
R(t) \simeq (\lambda_{2d} t)^{1/2}
\; . 
\end{equation}
As already said, the parameter $\lambda_{2d}$ is temperature, and material or model, dependent.

\subsection{Evolution of a single domain}
\label{subsec.disk}

The coarse-grained domain growth process with non-conserved order parameter dynamics is
described with a scalar field that follows a time-dependent 
Ginzburg-Landau equation~\cite{Bray94}. From this equation, in the absence of thermal 
fluctuations, Allen and Cahn obtained a 
generic law that relates the  local velocity of a point on an interface and the 
local mean curvature~\cite{AlCa79}
\begin{equation}
v = - \frac{\lambda}{2\pi} \kappa
\; , 
\label{eq:allen-cahn}
\end{equation}
with $\lambda$ a material-dependent parameter. The effect of temperature is usually 
incorporated in the prefactor~\cite{SaSaGr83,FiWe92,LaGrVi93}, $\lambda(T)$.

In two dimensions, the area enclosed by a circle evolves in time as
$\dot{A}=2\pi R \dot{R}$. 
Under curvature driven dynamics, the domain wall 
velocity, $v=\dot{R}$, is given by the Allen-Cahn law (\ref{eq:allen-cahn}). 
For the chosen geometry $\kappa = 1/R$ and 
the area of the disk decreases linearly in time, $\dot{A}=-\lambda_{2d}$, 
with a rate that is independent of $A$.

In three dimension, the volume of a sphere evolves in time as 
$\dot V = 4\pi R^2 \dot R$, the mean curvature is 
$\kappa=2/R$, and the time variation of the volume is no longer
independent of its size, $\dot{V}=-4\lambda_{3d} R$. In $3d$ one can follow  
the surface area of the sphere, $A=4\pi R^2$, and find $\dot A = -8\lambda_{3d}$, 
or the area of the equatorial slice, $A=\pi R^2$, and find $\dot A = - 2\lambda_{3d} \equiv  
- \lambda_{\scriptsize \rm sl}$.

We wish to check whether, and to what extent, these results remain valid on a cubic lattice
with single spin flip dynamics.
The fact that the area of an initial square or circular
droplet in the $2d$IM model with zero temperature 
Glauber dynamics decreases to zero linearly in time was proven 
in~\cite{SaSaGr83,KandelDomany90,ChayesSchonmannSwindle95}. A rigorous bound, compatible with 
this time-dependence, was derived in~\cite{Caputo-etal11,Lacoin13} for the $3d$IM with the same 
$T=0$ dynamics. In the rest of this section we analyze other initial states evolving at
non-vanishing  sub-critical temperature.

\subsubsection{Single disk/sphere}
\label{subsubsec:single-disk}

\begin{figure}[htb]
\vspace{5mm}
\includegraphics[width=4.2cm,angle=270]{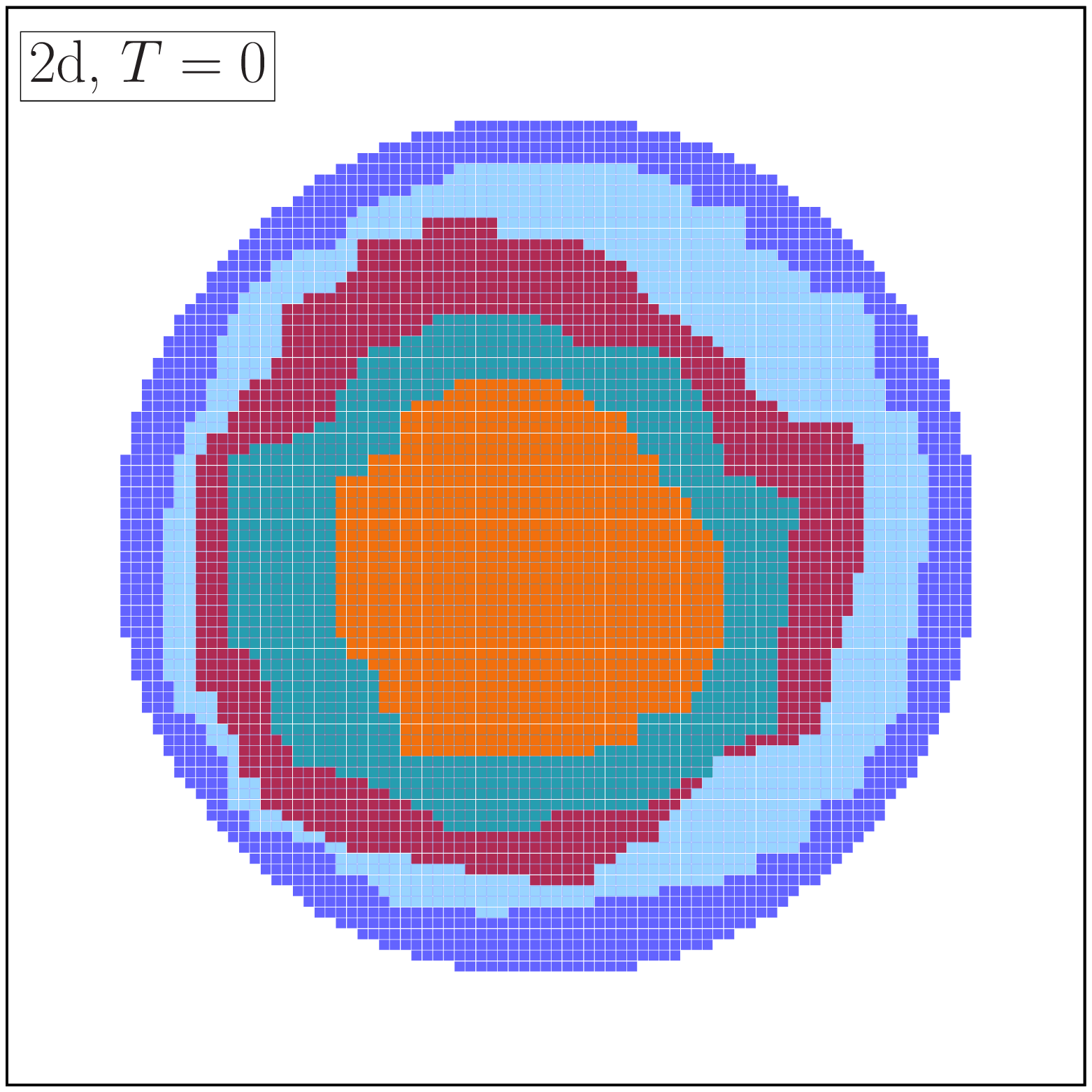}
\includegraphics[width=4.2cm,angle=270]{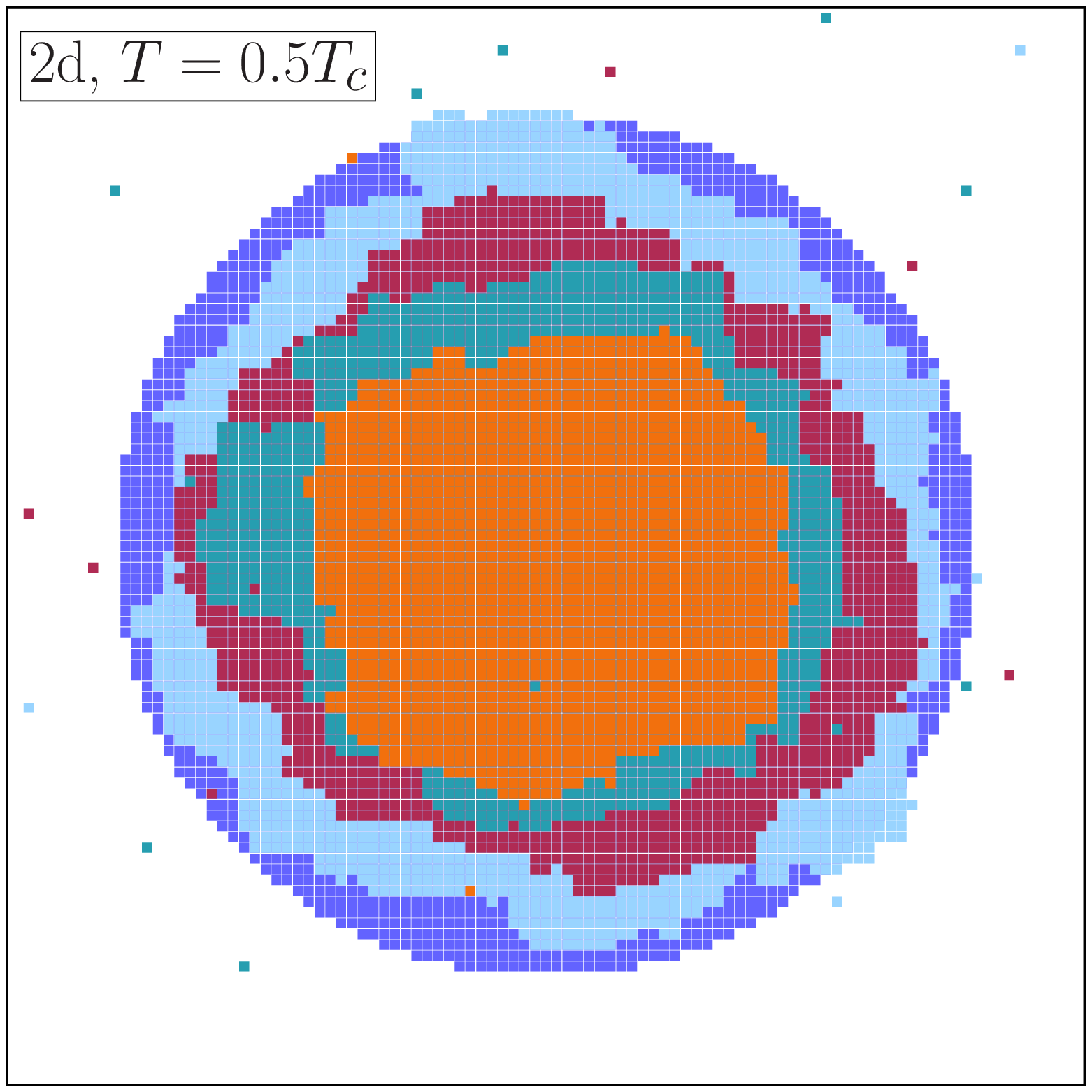}

\includegraphics[width=4.2cm,angle=270]{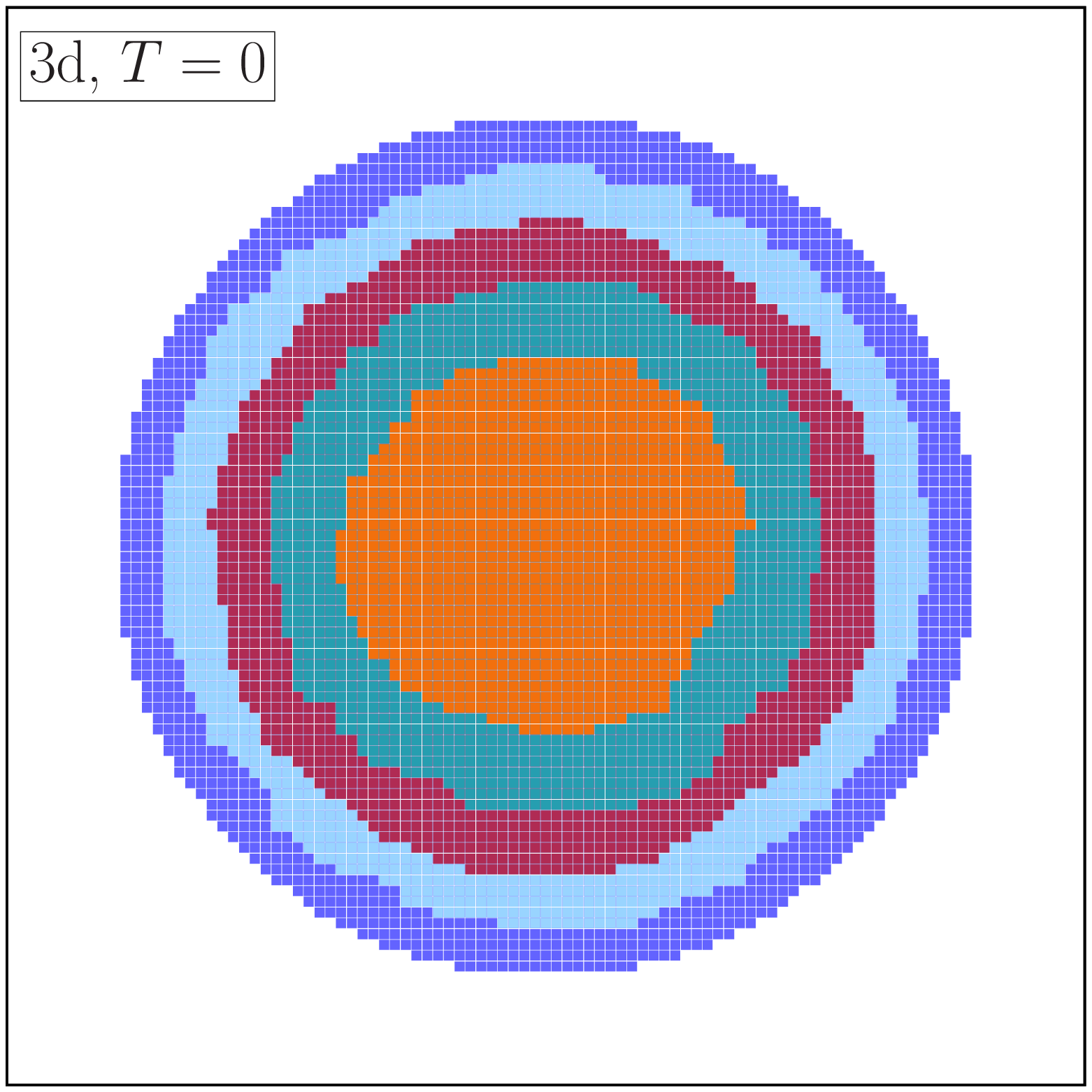}
\includegraphics[width=4.2cm,angle=270]{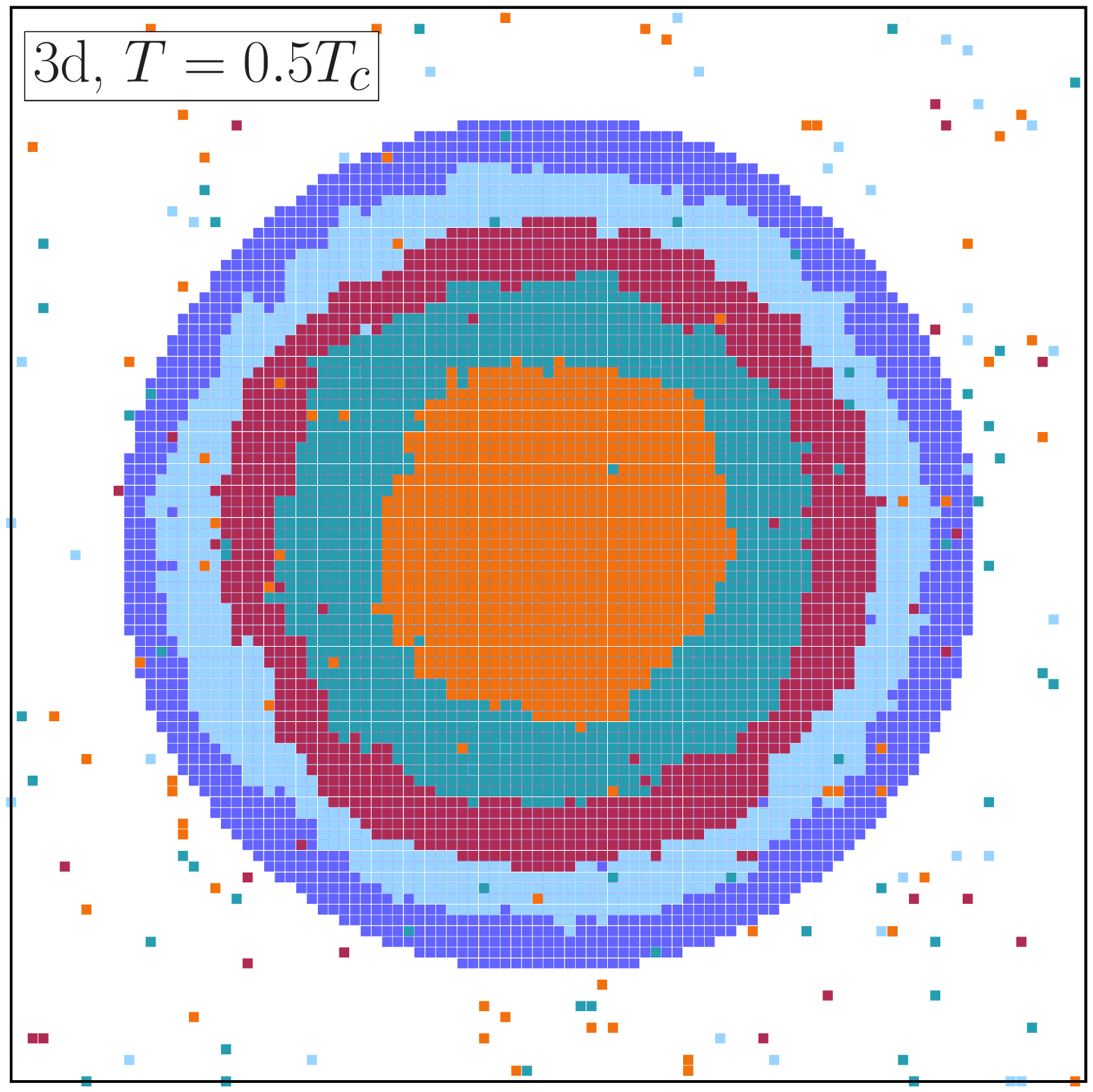}

\vspace{0.3cm}

\caption{(Color online.) Time evolution of a $2d$ circular domain (top row) and an equatorial
slice of a $3d$ spherical domain (bottom row) at $T=0$ (left column) and
$T=T_c/2$ (right column). In both cases the initial radius is $R_0=40$
(for finite temperatures, the box length must be large enough in order to 
prevent the circle from growing and percolating). 
Different colors correspond to different times
(in Monte Carlo steps, starting from the background: 0, 500,\ldots, 2500).
Although at zero temperature both areas decrease
at the same rate, $\lambda_{\scriptstyle\rm sl}(0)=\lambda_{2d}(0)\simeq 2$, in $3d$ the domains stay closer to their
circular initial shape at all times.}
\label{fig.snapshots.circle}
\end{figure}

Here we simulate the IM starting from a configuration in which all spins  that lie
inside a {\it circle} in $2d$ or a {\it spherical shell} in $3d$ point up, while all other spins point 
down. This configuration is then let evolve with MC single spin flip dynamics.
Figure~\ref{fig.snapshots.circle} shows some snapshots 
at different times, with and without temperature fluctuations. 

By measuring how the size of the original bubble changes in time from the data 
gathered at zero temperature and shown on the left column, one verifies that
the above relations for both $\dot{A}$ and $\dot{V}$ are satisfied at 
all times, with $\lambda_{2d}(0) \simeq 2$ (consistent with Ref.~\cite{SaSaGr83}) and $\lambda_{3d}(0) \simeq 1$, respectively. 
Notice that, with these values,  the product $\lambda\kappa$ is the same in $2d$ and $3d$
and $v=-(\pi R)^{-1}$. As a consequence, whatever the dimensionality, the 
radius behaves as 
\begin{equation}
R^2(t) = R_0^2 - \frac{2}{\pi}t
\; .
\end{equation} 
Therefore, an equatorial
slice of the $3d$ sphere and the $2d$ disk should show the same behavior. Indeed,
the area of the disk also decreases linearly in time with the same $T=0$
coefficient, $\lambda_{\scriptstyle\rm sl}(0)=2 \lambda_{3d}(0) = \lambda_{2d}(0)=2$. 

Interestingly, this value $\lambda_{\scriptsize \rm sl}=\lambda_{2d}=2$ is consistent with the average change,
$\langle \dot{A}\rangle$,  for a coarsening Ising 
system after having being quenched from an equilibrium state at either $T_0\to\infty$ or $T_0=T_c$ into
the low temperature phase, in which case the initial domains were no longer circular~\cite{LoArCuSi10,LoArCu12}.
We conjecture that the fitting value $\lambda_{2d}(0)\simeq 2.1$, obtained in 
Refs.~\cite{ArBrCuSi07,SiArBrCu07}, is indeed exactly 2.

Turning now the temperature on (data shown on the right column) we
checked that 
\begin{equation}
\lambda_{\scriptstyle\rm sl}(T)\simeq 2\lambda_{3d}(T)
\label{eq:slice-vs-3d}
\end{equation}
at all temperatures.

A striking difference between the $2d$ and the $3d$ cases, both at zero and at non-vanising 
temperature,  is that the surface of the sliced $3d$ system 
stays closer to its original circular shape at all times, while the $2d$ system
becomes more irregular. These surface fluctuations, stronger in $2d$, are much suppressed
in $3d$ because of the extra surface tension along the direction orthogonal to the slice.

\begin{figure}[htb]
\includegraphics[width=8cm]{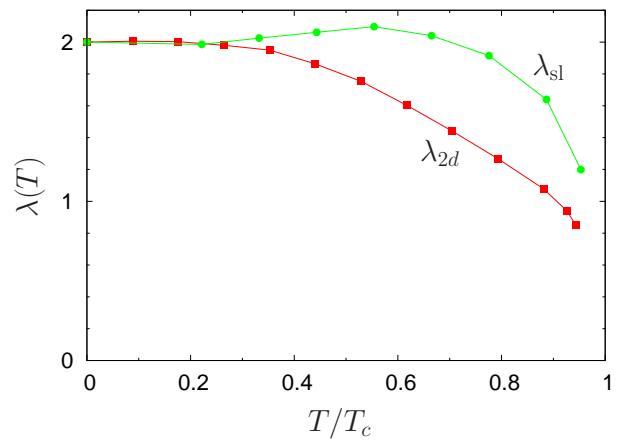}
\caption{(Color online.) The parameter $\lambda(T)$, obtained
from the shrinking of a single disk, $\lambda_{2d}$, or an equatorial slice of a single sphere, 
$\lambda_{\scriptsize \rm sl}$, as
a function of temperature.}
\label{fig.singlelambda}
\end{figure}

When the dynamics are affected by  thermal noise, the behavior of $\lambda$ depends
on the dimensionality, as can be seen in Fig.~\ref{fig.singlelambda}.  
Although $\lambda_{2d}(T)$, within our numerical precision, monotonically decreases as 
the temperature increases towards $T_c$~\cite{SaSaGr83,GrGu83}, this is not the case in $d=3$. 
Since the $3d$ system presents a large number of metastable states at 
$T=0$~\cite{SpKrRe02,OlKrRe11a,OlKrRe11b}, a small amount of noise may increase the 
wall velocity. Indeed, we find that $\lambda_{\scriptstyle\rm sl}(T)$
has a maximum at intermediate temperatures. Nevertheless, although the
temperature increases the roughness of the surface, 
the sliced disk still collapses more isotropically than the $2d$ one, the fluctuations
away from the circular shape being smaller.  
As the temperature approaches the critical value, $\lambda(T)$ tends to decrease to 
zero in both cases. It is, however, very hard to conclude about the exact $T$-dependence in this range
by tracking the evolution of a single initial volume. Some of the sources of 
difficulties are the fragmentation and merging processes that occur because of
the thermal fluctuations. Analogously, for the $3d$ case, isolated domains in the slice may
belong to the same three-dimensional cluster. 

 \subsubsection{Single toroidal domain}
 \label{subsubsec:single-torus}
 
 In the continuous description of $2d$ coarsening the areas evolve independently of each other.
Lattice effects do not affect this result at sufficiently large scales. However, although the 
dynamic mechanism in $2d$ slices of a $3d$ system is still curvature driven, the evolution of the 
areas on the slice may no longer be independent when, for instance, two different areas on a slice 
do belong to the same three dimensional domain. 

\begin{center}
\begin{figure}[h]
\includegraphics[width=8cm,angle=-90]{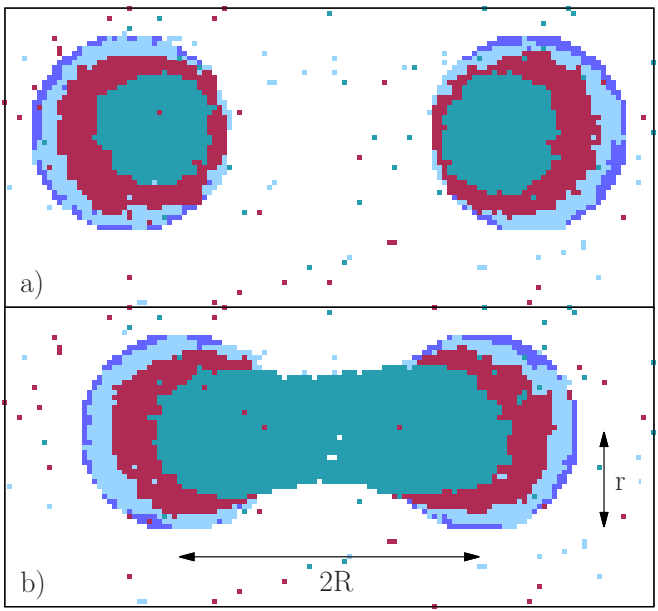}
\vspace{0.3cm}
\caption{(Color online.) Evolution of a toroidal domain of one phase immersed in a sea of the 
opposite phase at $T=T_c/2$  for two initial conditions. The snapshots of the cross-section of a single 
ring torus are shown with different  colors at times $t=0, 150, 500$ and 1000 MCs (from
back to front). 
Due to the shrinkage of the torus, there appears to be a small attraction
between the domains as their centers get slightly closer with time. 
In both cases the minor radius
is $r = 20$ and, depending on the value of the major radius, $R=40$ (a) and
$R = 30$ (b), the two initially separated circles may merge. This merging
mechanism, that slows down the change in area, is only present in
the slices of a $3d$ system because the domains are connected
along the orthogonal direction.}
\label{fig.snapshots.torus}
\end{figure}
\end{center}

A simple  initial configuration that illustrates the importance of the third dimension and the 
new mechanism that may arise on the slice is a toroidal structure. In Fig.~\ref{fig.snapshots.torus} 
we show the time evolution of an initial toroidal domain observed on a plane that contains
its axis of revolution, that is, the initial state has two circular domains whose radii are
the minor radius $r$ of the ring torus. The separation between their centers is twice the major
radius $R$. In the two cases shown in the figure, the minor radius is the same,
$r = 20$, but the major radius is different: $R=40$ in (a) and $R = 30$ in (b). The simulation is 
performed at $T_c/2$.
In both cases the whole toroid shrinks, and this can be seen as an effective attraction between
the disks as they move towards each other. However, in the second case, the two initial disks
change shape and, after some time, they merge and form an elongated domain in the 
plane. In the first case, the two disks do not merge on the observed timescale. Thus,
differently from the pure $2d$ case where such mechanism is absent, this merging
process decreases the number of domains, increases the average area and thus slows
down the rate at which the average area decreases.

\vspace{0.5cm}

Whether or not the results for $\lambda$, shown in this section for the evolution of a single domain,
transpose to the coarsening problem, will be analyzed below. Moreover, to what extent the above
merging mechanism has an important role in this case is an open problem.
 
\subsection{Coarsening slices}

When the initial state, instead of being prepared as a single sphere immersed in a
sea of opposite spins, is taken
from the equilibrium distribution at a given temperature above or at the critical point, 
much larger systems must be used in order to 
improve the statistics. Nonetheless, in $3d$ severe restrictions on 
the total size of the system are imposed. We here consider systems with linear size up 
to $L=400$ and finite size effects may still be important.

In $2d$, the Ising model can be quenched to $T=0$ and yet evolve for
a certain time before approaching either the ground state
or a stripe state~\cite{SpKrRe01,BlPi13}, time that in many cases is enough 
to study coarsening phenomena~\cite{ArBrCuSi07,SiArBrCu07}. In $d=3$, however,
the $T=0$ dynamics  get easily stuck in a sort of sponge state~\cite{SpKrRe02,CoLiZa08,OlKrRe11a,OlKrRe11b}. 
To avoid this halting of the configuration evolution, after the system is equilibrated either at $T_0\to\infty$
or $T_0=T_c\simeq 4.51$, the quench is performed to a finite working temperature, $T=2$, 
well below $T_p < T_c$. 

\subsubsection{$T_0\to\infty$}

\begin{figure}[htb]
\includegraphics[width=8cm]{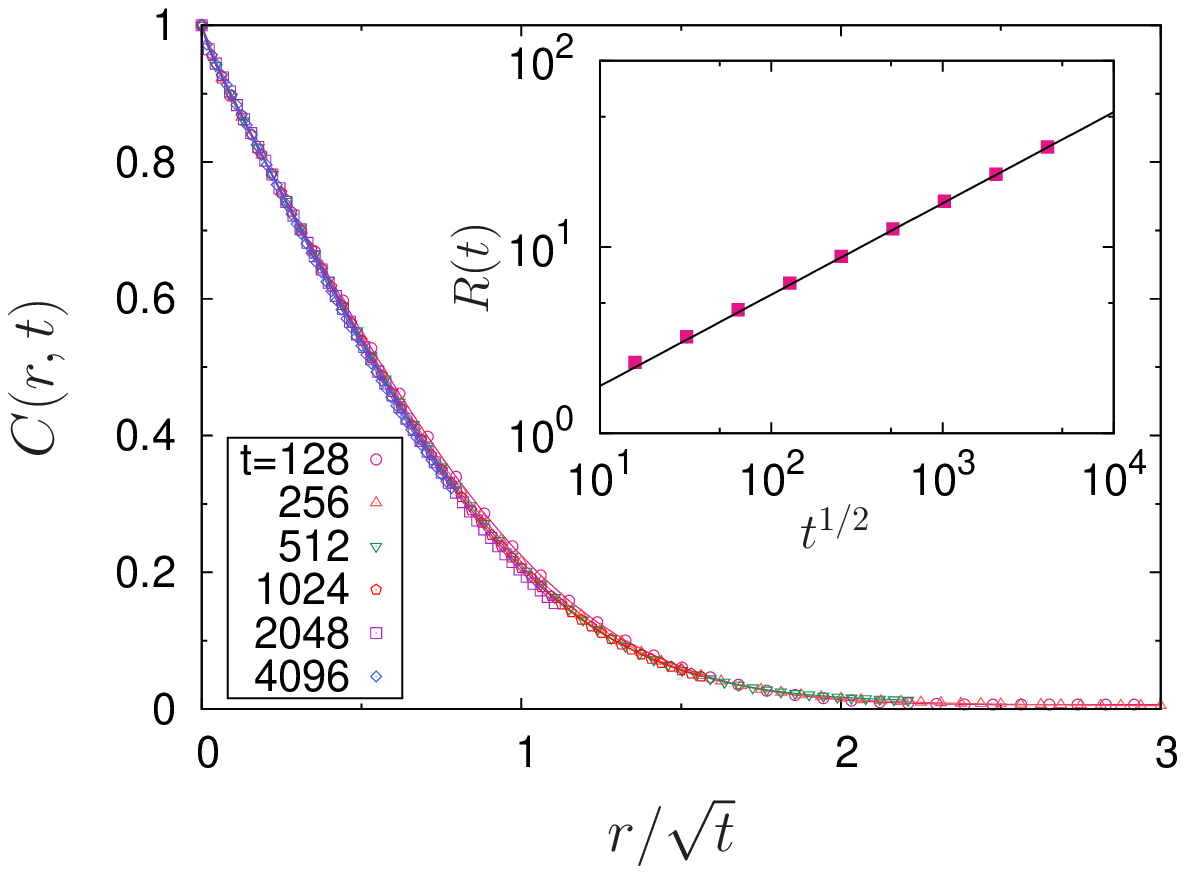}
\caption{(Color online.) Collapsed equal time correlation $C(r,t)$, for several 
times after the quench from $T_0\to\infty$, indicated in the key, as a function of the rescaled distance, $r/\sqrt{t}$.
As expected, dynamical scaling is observed. Inset: lenghtscale $R(t)$ obtained
from $C(R,t)=1/2$. The straight line has an exponent 0.5.
}
\label{fig.corr.slice.Tinf}
\end{figure}

We start the analysis by checking that dynamic scaling applies to correlation
functions measured on the slices in the usual way. In Fig.~\ref{fig.corr.slice.Tinf} we display the equal-time 
correlation between spins at a distance $r$ on the slice, $C(r,t)$, as a function of 
rescaled distance $r/t^{1/2}$, for several times given in the key. The scaling 
 is very satisfactory. In the inset we show  the evaluation of the growing length scale $R(t)$ 
using the criterium $C(R,t)=1/2$; the straight line is the $t^{1/2}$  growth law of curvature driven dynamics 
with non-conserved order parameter. Notice that although they could be taken into account,
we neglect the corrections to scaling linked to the time-scale $t_p$ discussed in Ref.~\cite{BlCoCuPi14} as
they are not necessary for our purposes here.
  
\begin{figure}[htb]
\includegraphics[width=8cm]{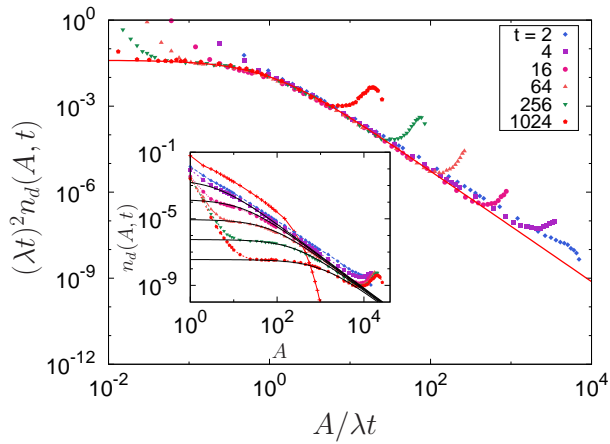}
\caption{(Color online.) 
Number density of geometric domain areas (inset) and its rescaled form (main panel), in a 
$2d$ slice of the $3d$IM (including the spanning clusters) per unit area of the system after a quench from $T_0\to\infty$ to $T=2$. 
Averages are over 7000 configurations (700 samples with 10 slices each)
of an $N=200^3$ system. Notice that since the linear system size is much smaller
than the ones used in Ref.~\cite{ArBrCuSi07}, the distributions have smaller cutoffs. 
Albeit the $3d$ system is far from the percolation threshold,
the distributions on the slice soon approach a power law distribution with an exponent that
is compatible, asymptotically and for large systems, with
$\tau_p=187/91\simeq 2.055$ (we use, indeed, the value obtained in Sec.~\ref{sec.eq} for
$L=200$: $\tau_A\simeq 1.93$). 
The lines are the $2d$ result, Eq.~(\ref{eq.analytic-nd}), with $c_d\simeq 0.02$, 
$\lambda_{3d}(2)\simeq 1.04$, obtained from the single sphere, and the slope $\tau_p$ above.
For small areas with respect to the typical one, $A/t<10$, we observe in the inset that the distribution of thermal fluctuations 
approaches its equilibrium form.} 
\label{fig.dist_nd_Tinf}
\end{figure}

At $T_0\to\infty$, the $2d$ slices of the $3d$ system are uncorrelated and any plane
is statistically equivalent to a pure $2d$ system. Since both species of spins have, on 
average, the same density, no domain percolates along the slices (on the square
lattice, $p_c \simeq 0.59$~\cite{Stauffer94}) and the distributions of areas and perimeters 
do not behave critically at $t=0$~\cite{ArBrCuSi07,SiArBrCu07} (see the corresponding 
curve in the inset of Fig.~\ref{fig.dist_nd_Tinf}). However, once quenched to a subcritical
temperature, the critical state of the $2d$ site percolation is approached
after a time that scales with the system size as $t_p\sim L^{\alpha_p}$.
The exponent $\alpha_p$ is 0.5 on the square lattice~\cite{BlCoCuPi14}
but we have not analyzed the scaling of $t_p$ for the $2d$ slices of the $3d$ system, 
what would constitute a project on its own. Nonetheless, since the phenomenology of 
both $d=2$ and $3d$ slices are similar (the area distribution soon develops a power 
law tail after the quench, as can be seen for $t=2$ and 4 in the inset of
Fig.~\ref{fig.dist_nd_Tinf}), we expect that $t_p$ will behave accordingly.
In the whole $3d$ volume, on the other hand, since the  
random site percolation threshold for the cubic lattice is $p_c=0.312$, there are percolating
clusters of both species of spins at $t=0$.
After the subcritical quench, the slices become 
correlated and the question we want to ask is to what extent the geometric properties, measured on a
slice, resemble those of a $2d$ system. 

\vspace{0.5cm}

\begin{figure}[htb]
\includegraphics[width=8cm]{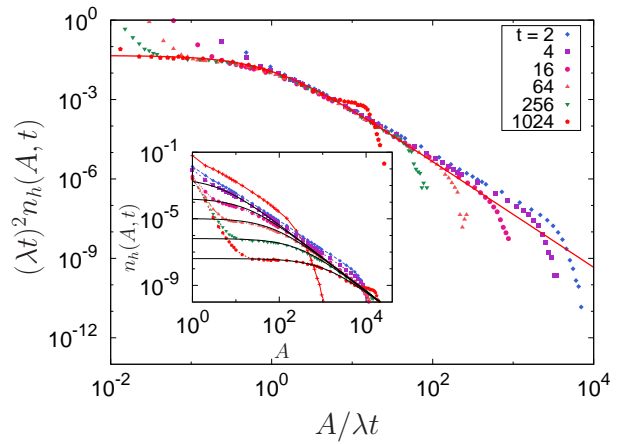}
\caption{(Color online.) 
The same as Fig.~\ref{fig.dist_nd_Tinf} but for the hull enclosed
areas. Differently from the geometric domains whose exponent $\tau_A$ has
strong size dependence, the power law exponent here is 2 and is attained
much faster. The lines are the $2d$ result, Eq.~(\ref{eq.analytic-nh}), with  
$\lambda_{3d}(2)\simeq 1.04$, obtained from the single sphere. Notice
that for long times, the distribution develops a bump, even though the
contribution from percolating clusters has been removed.} 
\label{fig.dist_nh_Tinf}
\end{figure}

\begin{figure}[htb]
\includegraphics[width=8cm]{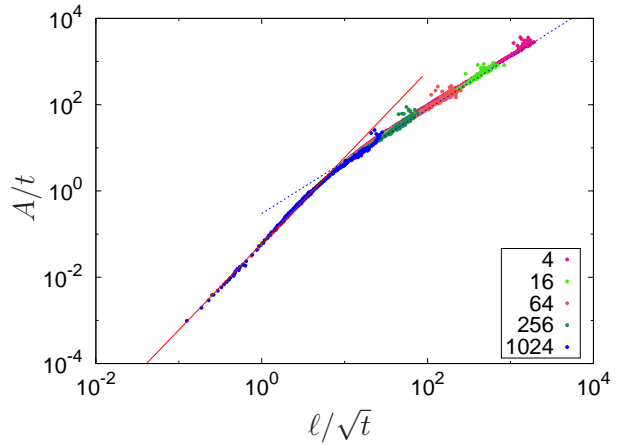}
\caption{(Color online.) 
Rescaled hull-enclosed areas against the corresponding rescaled perimeters for slices of a $3d$ system quenched
from $T_0\to\infty$ to $T=2$.  For small sizes with respect to the typical one, since domains get round, the rescaled areas are simply the square of the
rescaled lengths, $y\simeq x^2$. For large sizes, the rescaled quantities are linked by an exponent that is close to the 
one of critical $2d$ percolation, $8/7\simeq 1.14$~\cite{SaDu87,SiArBrCu07}. Numerically we find 
$y \simeq x^{1.2}$. Both behaviors, for small and large rescaled areas, are shown with straight dotted lines.
The times at which the data are gathered are shown in the key.}
\label{fig.apinf}
\end{figure}

As can be seen in both Figs.~\ref{fig.dist_nd_Tinf} and \ref{fig.dist_nh_Tinf}, the
exponent of the power-law tail increases with time. Its asymptotic value, for geometric domains
(Fig.~\ref{fig.dist_nd_Tinf}), is consistent with the critical percolation value, 
$\tau_p=187/91\simeq 2.055$~\cite{Stauffer94}, although determining it precisely is a very 
hard task, as already discussed in Sec.~\ref{sec.eq} for the equilibrium data. For hull
enclosed areas, Fig.~\ref{fig.dist_nh_Tinf}, on the other hand, the convergence to the
asymptotic exponent (2) is fast. The distributions for geometric domains contain all
clusters, even those percolating, and present an overshoot region that does not
change position as the system evolves (and thus moves to the left when we rescale the areas
by time, as shown in the main panel). For the hull enclosed areas there is no peak 
associated to the percolating domains (that are excluded by definition), but at later 
times the system develops a maximum anyway. The curves in these figures (inset) present, 
in the course of time, two other regimes. They all display a plateau, that crosses over to the 
power-law tail, and a first, very rapid decay at very small areas. The former is the actual curvature driven regime. The latter are static and due to equilibrium temperature fluctuations. 
In Fig.~\ref{fig.dist_nd_Tinf} (inset) the black solid lines represent the analytic 
law, Eq.~(\ref{eq.analytic-nd}), with $c_d\simeq 0.02$ and $\lambda(2) \simeq 1.04$. The constant $c_d$ takes the value 
used in Ref.~\cite{SiArBrCu07} for the $2d$ case. The factor $2$ in the numerator is associated 
to the high temperature initial condition. The parameter $\lambda$ is very close in value to the one measured for
the collapsing volume of a single 
sphere, see Sec.~\ref{subsec.disk}, evaluated at the working temperature $T=2$. 
Notice that even though the measurements are done on a slice, the relevant coefficient is the one obtained for the whole volume of the sphere. 
Analogously, in Fig.~\ref{fig.dist_nh_Tinf} (inset) the lines are Eq.~(\ref{eq.analytic-nh})
with the same coefficient $\lambda_{3d}$ and $c_h=1/(8\pi\sqrt{3})$, again closely following the $2d$ results~\cite{ArBrCuSi07}. Upon rescaling the areas by time, as
required by dynamic scale invariance, a rather good collapse of all curves onto a universal curve is found, see
Figs.~\ref{fig.dist_nd_Tinf}  and \ref{fig.dist_nh_Tinf} (main panels).  As time increases, the power law tail of the distributions is less visible (for these small system sizes). 

Areas and perimeters are also correlated~\cite{SiArBrCu07,LoArCu12}. As an example, 
we present the collapsed curves (rescaling the area $A$ by $t$ and the perimeter $\ell$ by $t^{1/2}$) in
Fig.~\ref{fig.apinf} for the hull-enclosed areas and the corresponding perimeters (for geometric 
domains, the perimeter would also include the internal perimeters). Small domains are compact
and round, thus $A\sim \ell^2$. Large domains are reminiscent of the large domains created soon 
after the quench, when the power law developed, and one expects that area and perimeter are
related as in critical percolation, $A\sim \ell^{8/7}$~\cite{SaDu87}. Numerically, we find an exponent 
close to 1.2, compatible with the $2d$ system value~\cite{SiArBrCu07,LoArCu12}, and with the results in
Ref.~\cite{DoPiWiHaMaMa95} for the equilibrium clusters. 
In summary, within the numerical precision of our simulation, the dynamical behavior on a slice 
of a $3d$ system is essentially equivalent to an actual $2d$ system when the initial
state is prepared at $T_0\to\infty$. The surprise is that instead of using the value of $\lambda$
obtained from the measurements on a slice of the $3d$ sphere, $\lambda_{\scriptstyle\rm sl}(T)$,
 the time evolving distribution of geometric domains uses $\lambda_{3d}(T)$
 related to the whole volume of the sphere, that is a half of the previous one, see Eq.~(\ref{eq:slice-vs-3d}).

\subsubsection{$T_0=T_c$}

\begin{figure}[htb]
\includegraphics[width=8cm]{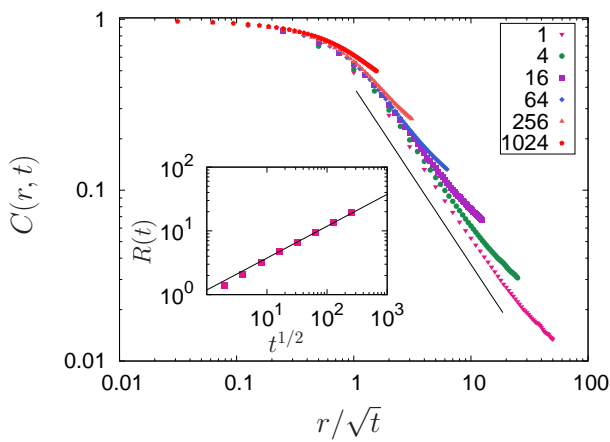}
\caption{(Color online.) Collapsed equal time correlation $C(r,t)$, for several 
times after the quench from $T_0=T_c$, indicated in the key, as a function of the rescaled distance, $r/\sqrt{t}$.
As expected, dynamical scaling is observed. Inset: lenghtscale $R(t)$ obtained
from $C(R,t)=1/2$. The straight line has an exponent 0.5.
}
\label{fig.corr.slice.Tc}
\end{figure}

As in the $T_0\to\infty$ case, we start the analysis by checking that dynamic scaling applies to correlation
functions measured on the slices also in the case in which the quench is performed from $T_0 = T_c$. 
We show in Fig.~\ref{fig.corr.slice.Tc} the equal-time 
correlation between spins at a distance $r$ on the slice, $C(r,t)$, as a function of 
rescaled distance $r/t^{1/2}$, for several times given in the key. Once again, the scaling 
 is good. Notice also that the decay of the correlation keeps memory of the power-law
present at the equilibrium state at $t=0$, $r^{2-d-\eta}$. Since the correlation is
isotropic, measuring $C(r,t)$ on a slice or in the whole volume would give the same
behavior, thus, in the power-law exponent, $d=3$ and $\eta=0.354$ (the
value for the $3d$ Ising model).
We present in the inset  the growing length scale $R(t)$ extracted from 
 $C(R,t)=1/2$ and the straight line  $t^{1/2}$.

\begin{figure}[htb]
\includegraphics[width=8cm]{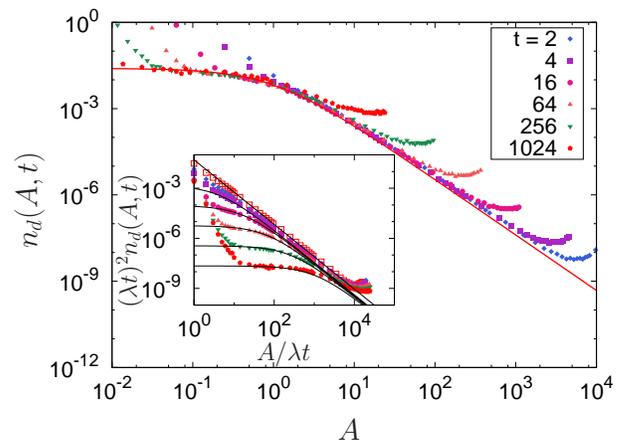}
\caption{(Color online.) 
Number density of geometric domains per unit area after a quench to $T=2$ 
in a $2d$ slice of a $3d$IM evolving from a
$T_0=T_c\simeq 4.5115$ initial condition. Averages are over more than 7000 configurations. 
The solid line on top of the $t=0$ data (inset, empty symbols) is $2c_d/A^{1.93}$ while for $t>0$
we use Eq.~(\ref{eq.analytic-nd}) without the factor 2 in the coefficient. (Main panel)
Collapsed version after properly rescaling both axes. The solid line has the same exponent
as the $t=0$ distribution but half the coefficient, $c_d/A^{1.93}$. Although the data could
be well enveloped by a power law with a slightly smaller exponent than 1.93, we remark
that at later times, due to the small size of the slice, the power law regime is barely
observed in the simulation.}
\label{fig.distg_Tc}
\end{figure}

When prepared in an equilibrium state at the critical temperature $T_0=T_c$, several geometric
distributions of the $2d$ system present power law behavior since, in this case, the
thermodynamical and the percolation transition coincide. Although this is no longer 
the case in $3d$ in which the percolation critical temperature associated with geometric
domains is smaller than the thermodynamical one, a $2d$ slice presents critical behavior at $T_c$ and, as
a consequence, one should find power-law behavior for several size distributions. 

We saw in Sec.~\ref{sec.eq} that these distributions present strong finite size effects
and, in particular, the exponents are smaller than expected. For example, the known exponent 
for the distribution of geometric domain areas in the critical  $2d$IM is  $\tau_A^{(2d)}=379/187$
but this value is only approached asymptotically, for very large system sizes. 
For smaller rescaled sizes, the apparent exponent is even smaller than 2 what would bring normalization issues. 
These effects are even 
stronger in a sliced $3d$ system, for which the data not even allow a clear 
extrapolation of the exponent. Besides differing in the behavior of the exponent, 
the coefficient of the power law distribution for a slice seems to have twice the 
value of the corresponding $2d$ distribution. It is thus interesting to see
how these differences occurring at $t=0$ evolve after the system is quenched
to a temperature lower than the critical one.

\vspace{0.5cm}

\begin{figure}[htb]
\includegraphics[width=8cm]{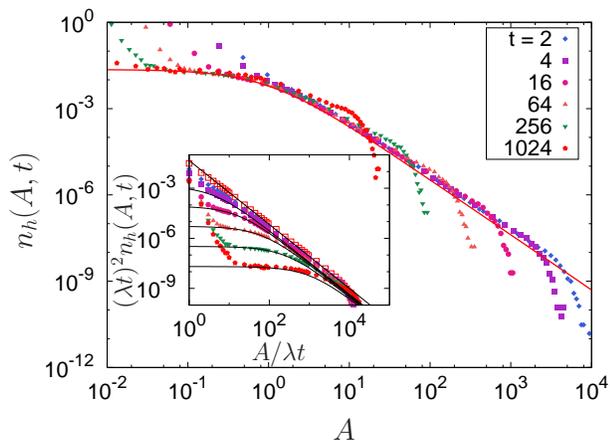}
\caption{(Color online.) The same as Fig.~\ref{fig.distg_Tc} but for hull enclosed areas.
The solid line on top of the $t=0$ data (inset, empty symbols) is $2c/A^{1.92}$ while for $t>0$
the distribution no longer has the factor 2 in the coefficient. The main panel
shows the collapsed version after properly rescaling both axis. Again,
the solid line has the same exponent as the $t=0$ distribution but half the coefficient, $c_d/A^{1.92}$.}
\label{fig.disthull_Tc}
\end{figure}

After the quench, there seems to be a very fast crossover to a distribution without
the extra factor 2 in the coefficient, see Fig.~\ref{fig.distg_Tc} (inset). This is
observed in the behavior of the distribution for small areas, as it approaches
a constant value that does not depend on the exponent, only on the coefficient and
the measuring time. Indeed, the curves in Fig.~\ref{fig.distg_Tc} (inset) 
are well fitted using Eq.~(\ref{eq.analytic-nd}) without the factor 2 in the coefficient.
However, for large areas, the tail of the distribution is not well described, as
one would expect, by Eq.~(\ref{eq.analytic-nd}) and the exponent measured at $t=0$, 
$\tau_A\simeq 1.93$. We must notice, however, that the slices are small, thus the
range of possible areas is rather limited. As time increases, the almost flat
part of the distribution gets wider and the power law regime is hardly 
observed. In addition, the system suffers from finite size
effects, even more severe than those for the $2d$ case as discussed in the
previous section, and the observed $\tau_A$ does not even extrapolate to the
right value. Nevertheless, we still observe the correct scaling as shown
in Fig.~\ref{fig.distg_Tc} (main panel). A similar behavior, but with an
exponent slightly smaller, is shown in Fig.~\ref{fig.disthull_Tc} for hull
enclosed areas.

\begin{figure}[htb]
\includegraphics[width=8cm]{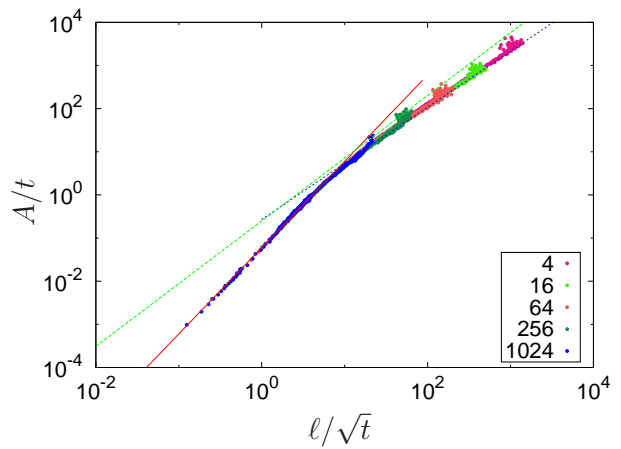}
\caption{(Color online.)
Rescaled hull-enclosed areas against the corresponding rescaled perimeters for slices of a $3d$ system quenched
from $T_0=T_c$ to $T=2$.  For small
rescaled domains $y\sim x^2$ (dotted red line) 
while for the larger ones $y\sim x^{1.3}$ (dotted straight blue line). This
exponent is compatible with the one for the 
geometric domains on $2d$ slices of the $3d$IM in equilibrium at $T_c$, 
$\delta \simeq 1.23$~\cite{DoPiWiHaMaMa95}, while it is 
well below the critical $2d$ exponent,  $\delta^{(2d)} \simeq 1.45$~\cite{Stauffer94}
(the dotted green line). 
}
\label{fig.apcrit}
\end{figure}

Areas and perimeters present, again, a two regimes relation. Small domains
are round and $A\sim \ell^2$. This first regime can be observed in the small
$A$ part of Fig.~\ref{fig.apcrit}, in which we related the size of a hull
with the area that it encloses. Larger domains, on the other hand, still
encode some information on its original shape and deviate from the
circular format. Indeed, roughly above $A/t\simeq 10$, the exponent
decreases to 1.3. This value is compatible with
previous estimates~\cite{DoPiWiHaMaMa95}, yet well below the critical
$2d$ exponent~\cite{VaSt89}, $16/11\simeq 1.454$. Again the origin for
such discrepancy may be the strong finite size effects previously
discussed.

\vspace{0.5cm}

\section{Conclusions}
\label{sec:conclusions}

We addressed the differences between the clusters of a $2d$ slice of a $3d$IM
and the ones of an actual $2d$ system, both in equilibrium and while coarsening. 

We recalled that the clusters on $2d$ slices of the $3d$IM {\it are critical} in equilibrium at $T_c$
(contrary to the $3d$ structures). We found that the distribution of finite clusters, $N(A)$, has a 
larger weight on the $2d$ slices than on the truly $2d$ model, while a second very large (though still 
finite) cluster is mostly absent in the former while present in the latter.  
We showed that although working with rather large system
sizes the measured exponents are still far from their asymptotic values when working on the 
$2d$ slices.

Next we moved to the analysis of the geometric clusters and hull-enclosed areas that develop 
after instantaneous quenches from equilibrium at the infinite and the critical temperature.
We found striking differences between the case with  long range correlations
in the initial state ($T_0=T_c$) and the case in which these do not exist ($T_0=\infty$). 
In the absence of correlations, neighboring layers
are independent, and even though strong correlations are built after a sudden
subcritical quench, the subsequent behavior does not essentially 
differ (within our numerical precision) from the one found in the strictly $2d$ case. 
On the other hand, for critical initial states, distant slices are correlated initially and such effect
introduces differences between properties of the slices and the actual
$2d$ system. These differences already exist in the initial state, as explained in the previous paragraph. 
The extra weight on the finite size areas (a factor 2) seems to be washed out 
very rapidly after the quench and the small rescaled areas on the $2d$ slices soon become very similar 
(identical within our numerical accuracy) to the ones of the $2d$ system.  
Instead, the distribution and geometric properties of the large objects are much harder to 
characterize numerically on the $2d$ slices as they are affected by strong finite size effects.
Although we find that the data satisfy dynamic scaling we cannot 
draw precise conclusions about the exponent characterizing the tail of the distribution
or the area-perimeter law as these are hard to determine numerically with good precision.

Work is in progress to extend these results to the $3d$IM with order
parameter conserving dynamics and to the Potts model.

\begin{acknowledgments}
JJA acknowledges the warm hospitality of the LPTHE (UPMC) in Paris 
during his stay where part of this work was done.
JJA is partially supported by the INCT-Sistemas Complexos and the
Brazilian agencies CNPq, CAPES and FAPERGS. LFC is a member of Institut Universitaire de 
France.

\end{acknowledgments}


\end{document}